**The U.S. Mortality Crisis as a Preston Curve Reversal**

Ritikaa Khanna (University of Pennsylvania)
Rourke O'Brien (Yale University)
Andrew C. Stokes (Boston University)
Atheendar S. Venkataramani (University of Pennsylvania and NBER)
Elizabeth Wrigley-Field (University of Minnesota)
(*Authors listed in alphabetical order*)

July 14, 2026

**Abstract:** U.S. life expectancy stagnated and declined in the 2010s despite continued growth in real per capita income. We use Preston curves to characterize this pattern as a change in the relationship between income and longevity. Using state-level data from 1980 to 2019 and county-level data from 2000 to 2019, we estimate population-weighted Preston curves relating life expectancy to logged real per capita income. From 1980 to 2010, U.S. states followed the classic Preston curve pattern: rising income was accompanied by rising life expectancy. Counties followed the same pattern from 2000 to 2010. From 2010 to 2019, however, states and counties continued to become richer while life expectancy stagnated or declined. The curves shifted right without shifting up and became steeper, indicating decoupling and divergence: increases in aggregate resources over time no longer produced broad longevity gains, and, in any given year, inequality in life expectancy by income grew. These patterns are robust to alternative temporal anchors around the Great Recession and to substituting education for income. They also appear across sex and racial groups. County-level decompositions are broadly consistent with arguments that longevity has fallen due to widely shared exposures to social deterioration, which may account for the Preston curve reversal. Collectively, we show that the recent U.S. mortality crisis reveals a weakening—and growing inequality—in the conversion of aggregate resources into longevity gains. We conclude that the recent U.S. mortality crisis could be understood not only as a story about particular causes of death, but also as a weakening of institutional and social translation.

**Notes**: We thank Michelle Niemann, Hannes Schwandt, Andrea Tilstra, Kristen Underhill, and participants at the Penn Center for Health Incentives and Behavioral Economics Roybal Retreat and Population Association of America 2026 Annual Meeting for helpful comments and suggestions. Elizabeth Wrigley-Field acknowledges funding from the Minnesota Population Center, funded through a grant from the Eunice Kennedy Shriver National Institute for Child Health and Human Development (NICHD; P2C HD041023) and the Russell Sage Foundation Visiting Scholars program.

## 1. Introduction

Across the twentieth century, rising economic resources and rising life expectancy moved together in the United States and across much of the world. But that relationship was never mechanical. Preston's (1975) classic analysis showed that income growth mattered partly because it coincided with the diffusion of medical knowledge, public health capacity, and mortality-reducing technologies. Societies became richer, but they also became better able to use their riches to prevent disease, reduce mortality risks, and extend longevity.

This historical pattern has become harder to reconcile with recent U.S. mortality trends. After decades of improvement, U.S. life expectancy slowed, stagnated, and then declined in the 2010s—even before the mortality shock of the COVID-19 pandemic (Ho and Hendi 2018; Woolf and Schoomaker 2019). Existing accounts have emphasized rising drug overdose mortality, suicide, alcohol-related mortality, stalled progress against cardiovascular disease, widening educational gradients, and increasingly divergent state policy contexts (Case and Deaton 2015; Ho and Hendi 2018; Montez et al. 2020; Woolf and Schoomaker 2019; Abrams et al 2023; Abrams et al 2026; Abrams and Brower 2026). This literature has substantially advanced understanding of the U.S. mortality crisis. Yet it leaves a broader demographic puzzle insufficiently characterized: why did a high-income country with rising aggregate resources stop producing broad gains in longevity?

We use Preston curves to bring this puzzle into sharper view. Preston curves are not merely descriptions of the cross-sectional association between income and life expectancy. They are tools for studying how aggregate resources are translated into population longevity. In Preston's original analysis, countries became longer-lived not simply because they became richer, but because mortality-reducing knowledge, technologies, and public health interventions diffused across populations (Preston 1975). The key issue, then, is not only whether richer populations live longer at a given moment, but whether rising resources continue to produce upward shifts in life expectancy over time.

Fundamental-cause theory helps explain why this translation may be unequal. Socioeconomic resources matter for health because they provide flexible means for avoiding risks and exploiting new opportunities for longer life (Link and Phelan 1995; Phelan, Link, and Tehranifar 2010). At the ecological level, richer places may likewise possess greater institutional capacity to adopt, finance, and distribute mortality-reducing protections. Subnational income per capita is therefore not only an indicator of individual resources. It also indexes collective capacity: tax bases, institutional resources, public infrastructure, educational and labor market conditions, and the ability of places to support health-protective environments. At the same time, states and counties are not separate societies. They are embedded within a national mortality regime shaped by federal policy, national markets, medical systems, technological diffusion, and shared institutional arrangements. The question is therefore how national and subnational institutions together mediate the relationship between rising aggregate resources and longer lives.

We argue that this relationship changed in the United States after 2010. Using state-level data from 1980 to 2019 and county-level data from 2000 to 2019, we show that U.S. Preston curves followed the expected historical pattern through 2010: places became richer and longer-lived. From 2010 to 2019, however, real per capita income continued to rise while life expectancy did not. The post-2010 pattern has two features. First, it reveals decoupling: the curve shifted right without shifting up, indicating that rising aggregate resources no longer produced broad longevity gains. Second, it reveals divergence: the curve steepened, indicating growing inequality across places in the association between income and longevity. The recent U.S. mortality crisis therefore reflects not only stalled life expectancy improvement but also widening geographic inequality in the conversion of collective resources into longer lives.

This study makes three contributions. First, it reframes the recent U.S. mortality crisis as a breakdown in the temporal movement of the Preston curve. The key fact is not simply that life expectancy stagnated or that places differ in mortality. It is that, since 2010, rising income stopped producing the broad upward shift in life expectancy that characterized earlier periods. Second, and consistent with other research on widening longevity gaps by income (Bor et al 2017) and education (Case and Deaton 2025; Hayward and Farina 2023; Chetty et al 2016), we show that this pattern is not an artifact of anchoring comparisons to 2010, a year close to the Great Recession. The same pattern appears when we use alternative pre- and post-recession temporal anchors and when we substitute education for income (the practice also advocated, for other reasons, in research by Lutz and Kebede (2018) that we engage with below). Third, we show that the shift is broad across major population groups and is not accounted for by standard socioeconomic, institutional, environmental, and demographic covariates.

Our aim is diagnostic rather than explanatory. We do not identify a single mechanism responsible for the post-2010 mortality shift. Instead, we identify several patterns that existing explanations must confront. The United States did not simply become poorer, less educated, or less insured in ways that mechanically explain the life expectancy shortfall. In many respects, observable aggregate resources improved. What changed was the relationship between those resources and longevity. The recent U.S. mortality crisis should therefore be understood not only as a story about particular causes of death, but also as a weakening of institutional and social translation: the failure of a rich society to convert available resources, flexible advantages, and mortality-reducing knowledge into broadly shared longevity gains.

## 2. Data and Methods

### 2.1. Data

We combine state- and county-level data on life expectancy, income, and socioeconomic conditions. State-level life expectancy estimates for 1980–2016 come from Woolf and Schoomaker (2019) and estimates for 2017–2019 come from the U.S. Centers for Disease Control and Prevention. County-level life expectancy estimates for 2000–2019 come from the Institute for Health Metrics and Evaluation (Institute for Health Metrics and Evaluation (IHME) 2022). State and county real per capita income comes from the Bureau of Economic Analysis (Bureau of Economic Analysis (BEA) 2026) and is adjusted to 2019 dollars.

The state analyses use observations for 1980, 1990, 2000, 2010, and 2019. The county analyses use 2000, 2010, and 2019, the decadal years (avoiding the Covid-19 pandemic) for which county-level life expectancy estimates are available. Together, these analyses allow us to compare the historical movement of the income–life expectancy relationship across several decades with the post-2010 period of life expectancy stagnation, and to assess whether the pattern appears at both state and county scales.

For stratified analyses, we estimate state-level Preston curves separately by sex and race, using estimates for female and male populations and for non-Hispanic Black and non-Hispanic White populations. These analyses assess whether the post-2010 shift appears broadly across major population groups rather than being confined to one subgroup.

For county-level decompositions, we focus on measures that reflect proximal demographic, institutional, and environmental factors that previous research has hypothesized to have contributed to shifting population health patterns in the 2010s. Economic measures include poverty rates, income inequality, college-educated shares, wealth, manufacturing employment shares, and labor force participation (e.g., Case and Deaton 2021; Machado et al 2025; Bor et al 2024; Deaton 2003; O'Brien et al 2022). Social and demographic measures include social capital, marriage, migration, ethnic fragmentation, and male labor force participation rates (Alesina et al 1999; Putnam 2001; Hochschild 2016; Metzl 2019). Our institutional measures focus primarily on government spending per capita and health insurance coverage (Huetsch et al 2026; Levy and Buchmueller 2025; Bradley et al 2016). We focus on air quality to capture environmental conditions (Manisalidis et al 2020).

We use these covariates to assess whether changes in observable county characteristics can statistically account for the observed life expectancy shortfall. We treat the results of these decompositions as descriptive facts with which any causal accounts of the Preston curve reversal must be consistent, and we comment at the end of this note on which accounts seem easiest or most difficult to reconcile with these results.

### 2.2. Methods: Preston Curves

We estimate Preston curves by modeling life expectancy as a function of real per capita income. The curves describe the ecological relationship between aggregate resources and population health. Our focus is not only the cross-sectional association in each year, but the movement of the curve over time. In the conventional Preston-curve pattern, rising income is accompanied by an upward shift in life expectancy. A rightward shift without an upward shift therefore indicates a weakening in the temporal translation of aggregate resources into longer lives. We also constructed the Preston Curves by plotting life expectancy against logged per capita income (Appendix Figure A2) and using a logged x-axis scale while retaining real per capita income (Appendix Figure A3). However, neither of the alternative approaches made any material changes to the results.

In the main specification, curves are weighted by population. We use a second-order polynomial in real per capita income to allow the income–life expectancy relationship to be nonlinear.

We interpret two features of the curves. The first is vertical movement: upward movement indicates that places with comparable income levels achieved higher life expectancy in later years. The second is slope: a steeper curve indicates greater inequality across places in the association between aggregate income and longevity. Together, these features allow us to assess whether rising resources were accompanied by broad longevity gains and whether those gains were shared evenly across places.

Because 2010 is shortly after the Great Recession, we test whether the post-2010 pattern is sensitive to the temporal anchor. The specific concern we want to alleviate is that the Preston curve might artifactually move downwards when compared with a 2010 baseline because 2010 incomes are systematically lower than a county's longer-term average income due to proximity to the Great Recession's period shock, while longer-run income might better reflect economic resources relevant to longevity in the way that the Preston curve is meant to capture. We estimate alternative state-level curves using 2006 life expectancy and 2006 real per capita income in place of 2010 values. We also pair 2012 life expectancy with 2006 real per capita income and 2010 life expectancy with 2010 proportion of the population with a Bachelor's degree (which we do not expect to be as sensitive to the Great Recession as 2010 income) (Figure 2). These checks do not estimate effects of the Great Recession; they assess whether the central Preston-curve pattern is robust to reasonable alternative baselines. In addition, we also assess whether Preston-curve patterns hold when using income per capita estimates adjusted for state- and region-specific price deflators, which accounts for the possibility that rising costs of living may contribute to shifts over time.

### 2.3. Methods: Socioeconomic Decomposition

Finally, we conduct descriptive county-level decompositions (Figure 4) to assess whether changes in a wide range of observed county characteristics can statistically account for the post-2010 life expectancy shortfall. These analyses are motivated by the possibility that income alone may not capture the relevant ecological determinants of life expectancy.

The decomposition follows the approach Lutz and Kebede (2018) apply to decompositions of education-based Preston curves. We estimate the relationship between county life expectancy and each of the characteristics in a baseline year, apply that relationship to later-year covariate values, and compare expected life expectancy with observed life expectancy in the later year. For example, we can estimate the relationship between life expectancy and labor force participation for the year 2010 and ask what life expectancy would be in 2019 given labor force participation in 2019 if the relationship between the two variables remained the same as in 2010.

We use this approach to estimate three comparisons: 2019 life expectancy predicted from 2010 relationships and 2019 covariate values; 2010 life expectancy predicted from 2000 relationships and 2010 covariate values; and 2019 life expectancy predicted from 2000 relationships and 2019 covariate values. These comparisons allow us to evaluate whether the 2010s were unusual relative to earlier covariate–life expectancy relationships and whether the 2000s and 2010s differed in how observed covariate change translated into longevity.

These decompositions do not identify the causal mechanisms responsible for the post-2010 mortality shift. Instead, they constrain the development of causal accounts by asking whether the observed life expectancy shortfall can be statistically accounted for by changes in standard socioeconomic, institutional, environmental, and demographic covariates. If observed life expectancy falls below expected life expectancy, then covariate change would have predicted greater longevity gains than occurred. If observed life expectancy exceeds expected life expectancy, then longevity improved more than covariate change alone would have predicted.

## 3. Results

### 3.1. Rising Income Without Rising Longevity

**Figure 1** shows Preston curves relating life expectancy to logged real per capita income across U.S. states and counties. Among states, the curves follow the classic Preston-curve pattern from 1980 through 2010: states became richer and longer-lived. The curve shifted right as real per capita income increased and shifted upward as life expectancy improved. Counties show the same pattern between 2000 and 2010. At both geographic scales, rising aggregate resources were accompanied by broad gains in longevity.

That pattern changed after 2010. Between 2010 and 2019, states and counties continued to become richer, but the curves did not shift upward. Life expectancy stagnated or declined despite continued increases in real per capita income. This is the central empirical fact revealed by Figure 1: resource accumulation continued, but commensurate longevity gains did not.

The post-2010 shift is also distributional. The curves became steeper, indicating greater inequality across places in the relationship between income and life expectancy. Higher-income states and counties were better able to preserve or extend longevity gains, whereas lower-income places experienced more pronounced stagnation or decline. The post-2010 period therefore involved both decoupling and divergence: rising income no longer translated into broad longevity improvement, and places became more unequal in their capacity to convert aggregate resources into longer lives.

### 3.2. The Pattern Is Not an Artifact of Recession Anchoring Or Regional Differences in Inflation

A possible concern is that the 2010 curve partly reflects the unusual economic and mortality conditions surrounding the Great Recession. Recessions can affect both income and mortality, complicating comparisons anchored to recession or immediate post-recession years. **Figure 2** addresses this concern by using alternative temporal anchors.

The results are similar when 2006 life expectancy and income are used instead of 2010 values. They are also similar when 2012 life expectancy is paired with 2006 income. In both specifications, the post-2010 pattern remains: real per capita income increased, but life expectancy did not rise in the way observed in earlier decades. The decoupling shown in Figure 1 is therefore not an artifact of using 2010 as the comparison year or of measuring income during the Great Recession.

**Figure 2** supports the interpretation that the post-2010 shift represents a real change in the temporal movement of the income–longevity relationship. Even when the comparison is anchored before the Great Recession, the United States became richer without achieving the broad upward shift in life expectancy that characterized earlier decades.

Appendix Figure A4 plots Preston curves by measures of real per capita adjusted for regional price deflators. The pattern we find is nearly identical to Figure 1, suggesting that differences in cost of living are unlikely to fully account for the post-2010 shift in the Preston curve.

### 3.3. The Shift Appears Across Population Subgroups

**Figure 3** shows state-level Preston curves stratified by sex and race. The purpose of this figure is not to emphasize differences across groups, although those differences are substantial and well known. Women have higher life expectancy than men, and White Americans have higher life expectancy than Black Americans. The key point is the commonality of the temporal pattern. For women and men, and for Black and White populations, the curves shift upward before 2010 and then stagnate or decline after 2010.

This commonality matters. The post-2010 shift is not confined to one major demographic subgroup. It appears among women as well as men and among Black Americans as well as White Americans. The aggregate pattern is therefore not a compositional artifact produced by deterioration in a single population group. It reflects a broader change in the relationship between place-level income and longevity.

Appendix Figure A1 extends this analysis using age-specific all-cause mortality rates. Those results show that the post-2010 shift is especially visible among younger and working-age groups, while older ages exhibit more conventional mortality improvement. The age-specific patterns are consistent with the importance of midlife mortality in recent U.S. trends, but they also show that the Preston-curve shift is broader than a narrow deaths-of-despair account. The phenomenon is better understood as a broad change in the U.S. mortality regime, with especially pronounced deterioration below older ages.

### 3.4. Socioeconomic Change Alone Does Not Account for the Post-2010 Shortfall

**Figure 4** compares observed life expectancy with life expectancy expected from earlier relationships between life expectancy and socioeconomic, institutional, environmental, and demographic covariates. The figure asks a simple counterfactual question: if earlier covariate–life expectancy relationships had persisted, how much life expectancy would later years have achieved, given the actual changes in these covariates?

Panel A predicts 2019 life expectancy using relationships estimated in 2010. For many socioeconomic covariates, observed 2019 life expectancy falls below expected life expectancy. Changes in income per capita, college-educated share, wealth per capita, and health insurance coverage would have predicted higher life expectancy than the United States actually achieved in 2019. The post-2010 shortfall therefore cannot be accounted for by changes in these socioeconomic characteristics alone. On the other hand, life expectancy in 2019 was higher than what was predicted by changes in poverty rates, income inequality, labor force participation, and sociodemographic measures like marriage rates, migration rates, social capital, and ethnic fragmentation. One interpretation of this Panel is that  broad measures of social breakdown and economic inequality acted to slow down life expectancy improvements that would otherwise have accrued from improving overall economic measures.

Panels B and C place the 2010s in longer perspective by using relationships estimated in 2000. Observed 2010 life expectancy exceeds what would have been expected from many 2000 covariate relationships, indicating that the 2000s were a period of stronger-than-expected mortality improvement.

Together, the decomposition results sharpen the central interpretation. The 2010s mortality crisis did not occur because income, education, wealth, or insurance coverage simply moved in the wrong direction. In many respects, aggregate socioeconomic conditions improved. Instead, what changed was the relationship between those conditions and longevity. The United States became less successful at converting collective resources s into population-level longevity gains.

### 3.5. Summary

Collectively, these results establish four patterns that theories of life expectancy change will need to contend with. First, state and county Preston curves show that the historical upward shift linking rising income to rising life expectancy stalled after 2010. Second, this result is robust to alternative temporal anchors and also applies to college education shares. Third, these patterns appear across sex and racial groups and are consistent across  all-cause mortality patterns for multiple age groups (Figure A1). Fourth, county-level decomposition analyses indicate that standard socioeconomic covariates do not account for the post-2010 life expectancy shortfall.

The results do not identify a single mechanism responsible for recent U.S. mortality trends. Instead, they define the puzzle more sharply. The United States continued to accumulate aggregate resources, but those resources no longer translated into broad longevity gains. Any adequate account of the recent U.S. mortality crisis must explain this decoupling and this divergence.

## 4. Discussion and Conclusion

This study uses Preston curves to characterize a central feature of recent U.S. mortality trends. Through 2010, rising real per capita income was accompanied by rising life expectancy across U.S. states and counties. After 2010, however, the pattern changed in two ways.

The first is decoupling: after 2010, rising income no longer produced a general upward shift in life expectancy. The second –– consistent with other recent research (Chetty 2016; Currie and Schwandt 2020; Bor and Galea et al 2017; Montez et al 2020) –– is divergence: the income–longevity curve steepened, meaning that places became more unequal in the mortality returns to aggregate resources. The recent U.S. mortality crisis was not simply a national slowdown distributed evenly across the country. It was a slowdown accompanied by widening geographic inequality in the conversion of resources into longer lives.

This interpretation connects the mortality crisis to broader trends in inequality, but not only to inequality in the distribution of income or wealth. The results point to inequality in the conversion function linking resources to longevity. Places are unequal not only in how much income they have, but also in how effectively available resources are converted into longer lives. This is the place-level analogue of a core insight from fundamental-cause theory: flexible resources matter because they shape who can avoid risks and take advantage of new opportunities for longer life (although see Freese and Lutfey [2011] for an argument that fundamental causes should be understood more broadly than through resources). In the post-2010 United States, those mortality returns to aggregate resources appear to have weakened overall and become more unequal across places.

This framing complements recent research on U.S. mortality divergence and state policy contexts. Montez and colleagues have shown that state policy environments play an increasingly important role in shaping U.S. life expectancy and working-age mortality (Montez et al. 2020; Montez et al. 2022). Our findings provide a Preston-curve representation of that broader divergence: states and counties are increasingly unequal not only in longevity, but in the relationship between aggregate resources and longevity. State policy contexts are likely one important mechanism through which available resources either do or do not become longer lives. At the same time, states and counties remain embedded in a national mortality regime shaped by federal policy, national markets, medical systems, and shared institutional arrangements. The patterns documented here therefore point to failures of translation operating across multiple scales.

The results also return to a central insight of Preston's original work. Income matters for mortality, but changes in life expectancy over time are not reducible to income growth alone (Preston 1975). Mortality decline depends on the diffusion and distribution of health-protective knowledge, technologies, institutions, and policies. Fundamental-cause theory makes the same point in a different register: resources matter because they provide flexible means for avoiding risks and taking advantage of new opportunities for longer life (Link and Phelan 1995; Phelan, Link, and Tehranifar 2010). The post-2010 United States is troubling because aggregate resources continued to grow while longevity gains stalled. The issue is not the absence of resources, but the failure to convert available resources into broadly shared longevity gains.

The decomposition analyses reinforce this interpretation. Observed socioeconomic conditions did not simply deteriorate in ways that mechanically explain the post-2010 shortfall. In many respects, counties became richer, more educated, wealthier, and more insured. Earlier relationships between these covariates and life expectancy would have predicted greater improvement than actually occurred. Conversely, the 2000s appear to have been a period in which life expectancy improved more than many covariate relationships would have predicted. This contrast suggests that the recent mortality crisis cannot be understood only by asking which socioeconomic indicators changed. It also requires asking whether the institutions that translate social and economic resources into longer lives changed. The fact that life expectancy outpaced expected gains from measures of social well-being – like poverty, marriage rates, migration rates and social capital – suggests that broader social and institutional breakdown may have prevented the U.S. from meeting its true population health potential.

Several limitations are important. The analyses are ecological and descriptive. They do not identify individual-level mechanisms or estimate causal effects of income on mortality. State and county income measures broad aggregate resources and may proxy many compositional and institutional characteristics. The decomposition analyses are not causal identifications; they assess whether observed covariate change can account for the life expectancy shortfall under earlier covariate–life expectancy relationships. In that sense, they represent a thought experiment—what longevity would have resulted if covariate-longevity relationships had not changed since 2010?—that can guide the development of further hypotheses.

This paper does not adjudicate among cause-specific explanations for recent mortality trends. Instead, it places those explanations within a broader demographic pattern they must help explain. We propose four sets of candidate explanations for future research. First, per capita income may no longer reflect the social and material resources available to improve health. For example, concerns about affordability underscore the possibility that the same level of real income cannot purchase the same health-enhancing goods and services as it might have in previous decades. Our similar findings when using measures of income that address regional differences in cost of living alleviate this concern to an extent, but not fully. The shift in the responsibility of insuring against general risks and provision of caregiving and other social services to individuals and families may place greater demands on a given level of income, leaving less resources to support health (Calarco 2024; Hacker 2006). And ecological-level economic resources may increasingly be captured by a narrow set of actors who put them to non-productive ends (Del Río 2021; Mazzucato 2018; Zemaitis 2026; Melo and Miller 2022), limiting their impact on longevity.

A second potential explanation is generalized social alienation and breakdown, which has been documented in previous decades, but may have intensified since the Great Recession (Antonoplis et al 2026). Falling trust in institutions and rising disconnectedness and polarization may have uniquely affected health since 2010, in ways that have decoupled the typical income-health relationship. Consistent with this possibility is our finding that predicted life expectancies based on changes in measures of societal well-being are consistently lower than actual life expectancies – a sign that these factors may be holding back population health progress that otherwise would have occurred.

Third, a strand of emerging work suggests that the post-2010 phenomenon may be explained by cohort-based, rather than period-based, explanations (Reynolds 2025; Reynolds 2024; Abrams et al. 2026; Andrade et al 2025; Pifarré i Arolas et al., 2026). This body of work deserves serious consideration. Our analysis of Preston curves for age-specific mortality suggests a more generalized decline occurring in 2010, suggesting that (consistent with Abrams et al. 2026) period effects are playing an important role. That said, we do find particularly dramatic Preston curve reversals at young adult and midlife ages, which is partially, but not fully, consistent with the Reynolds (2024, 2025) account. That account would predict a stronger reversal at midlife ages rather than older ages, which is indeed what we find, but Reynolds's (2024, 2025) preferred explanation of early-life lead exposure is harder to reconcile with such a reversal for cohorts whose early lives occurred after major sources of lead exposure were substantially mitigated (McFarland et al. 2022).

Fourth, the post-2010 Preston curve reversal may in part reflect a reversion to longer-running trends in the income-life expectancy relationship. The late 1990s and early 2000s were characterized by improvements in medical technology for the management of cardiovascular disease (CVD) that made CVD events more survivable (Mehta et al 2010; Singh et al 2019; Abrams and Bower 2026). These technological changes may have quickened the pace of life expectancy improvements, particularly for older age groups and those who have the means to access new technologies, during this period. It is possible that the period in which U.S. counties and states displayed a clear Preston curve pattern are the truly anomalous ones in recent American history.

Finally, the broader implication of our study is that the U.S. mortality crisis is not only a crisis of particular causes of death. It also appears to be crisis in the social organization of longevity. The United States became richer, but the mortality returns to rising resources weakened and became more unequal across places. We do not explain why this translation function weakened. We show that it did. Future research should therefore ask not only why overdose, suicide, alcohol-related mortality, or cardiovascular mortality changed, but why available resources, flexible advantages, and mortality-reducing knowledge were not converted into broadly shared longevity gains — and why places with fewer resources fell further behind.

**Figure 1. Preston Curves of Life Expectancy and Real Per Capita Income (Logged) by Decade, U.S. States and Counties**

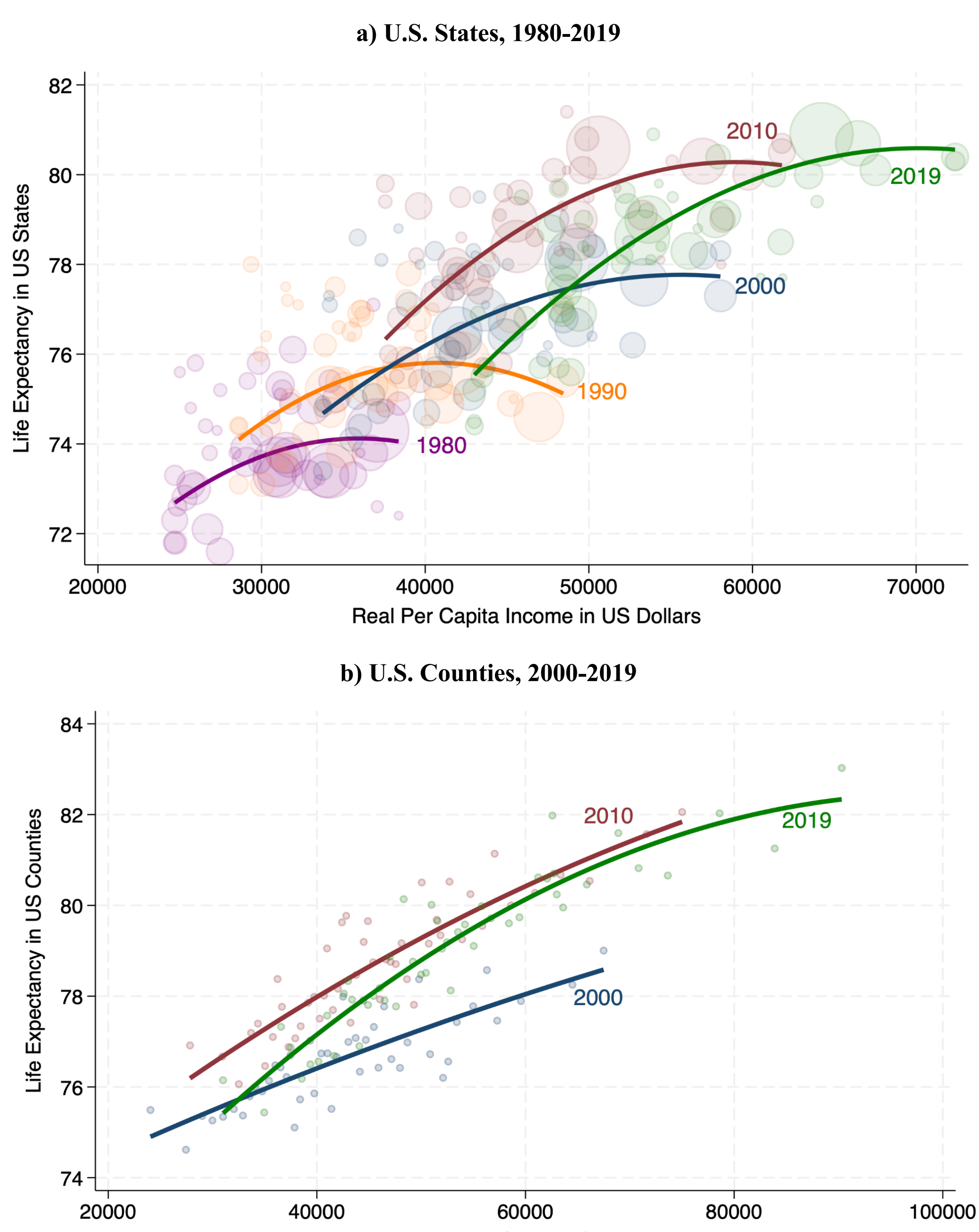

Notes: Preston curves for 1980, 1990, 2000, 2010, and 2019 (state) and 2000, 2010, and 2019 (county). For reach year, we regressed life expectancy on a quadratic of income per capita, and plotted the best fit lines. Regressions were weighted by state or county population (unweighted estimates were similar).

## Figure 2: Preston Curves Using Alternative Baseline Years and Income Alternative

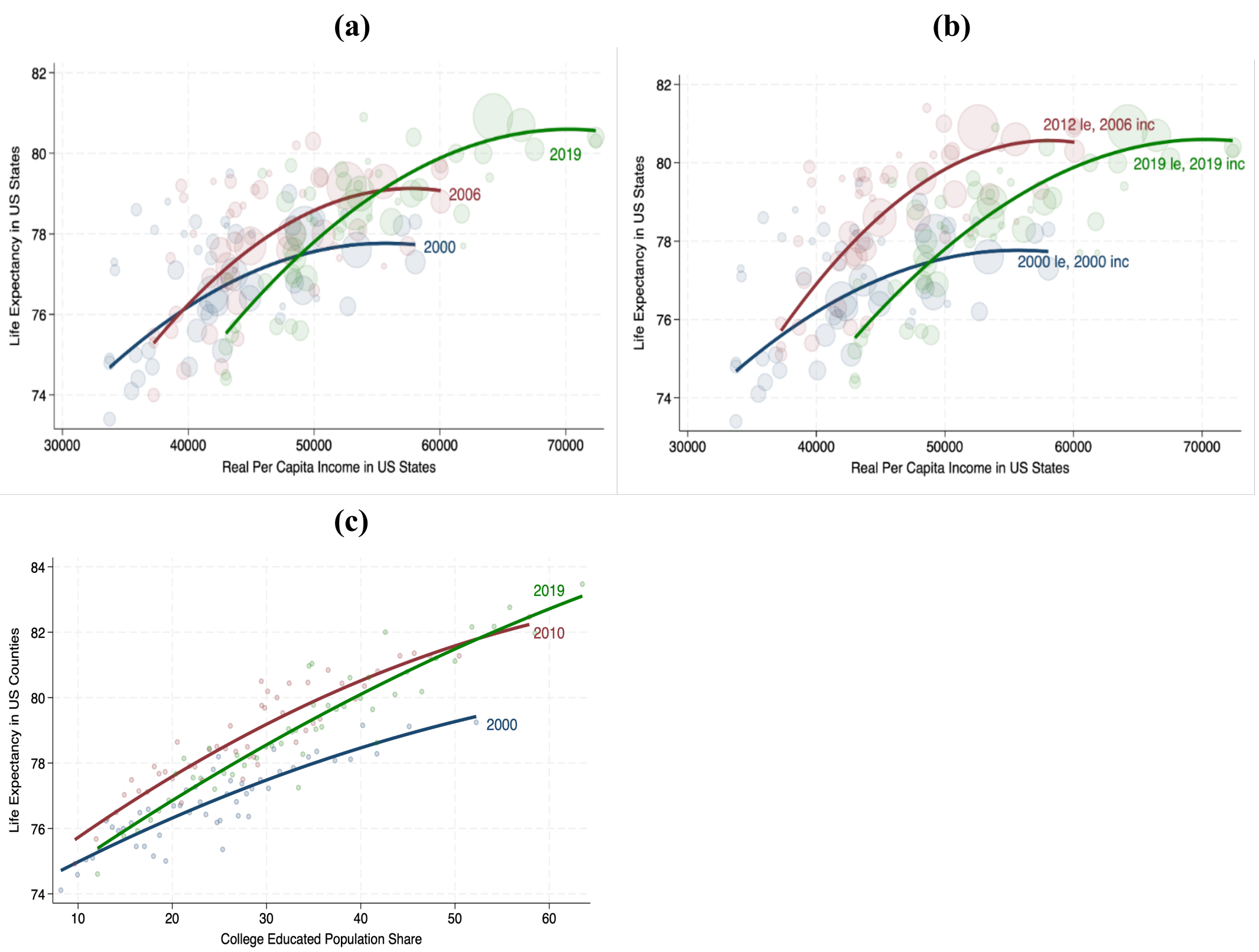

Notes: To assess whether estimates of the 2010 Preston Curve reflected transient shifts in income and health due to the Great Recession, we reconstructed curves for that year using baseline per capita income of 2006 (state level, panel a); 2006 income and 2012 estimates of life expectancy (state level, panel b); and college educated population shares instead of income per capita measures (county level, panel c). The overall shape and conclusions of the Preston Curves in all three alternatives remain unchanged. See Figure 1 for details on construction.

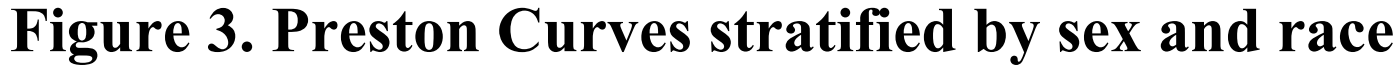

## Figure 3. Preston Curves stratified by sex and race

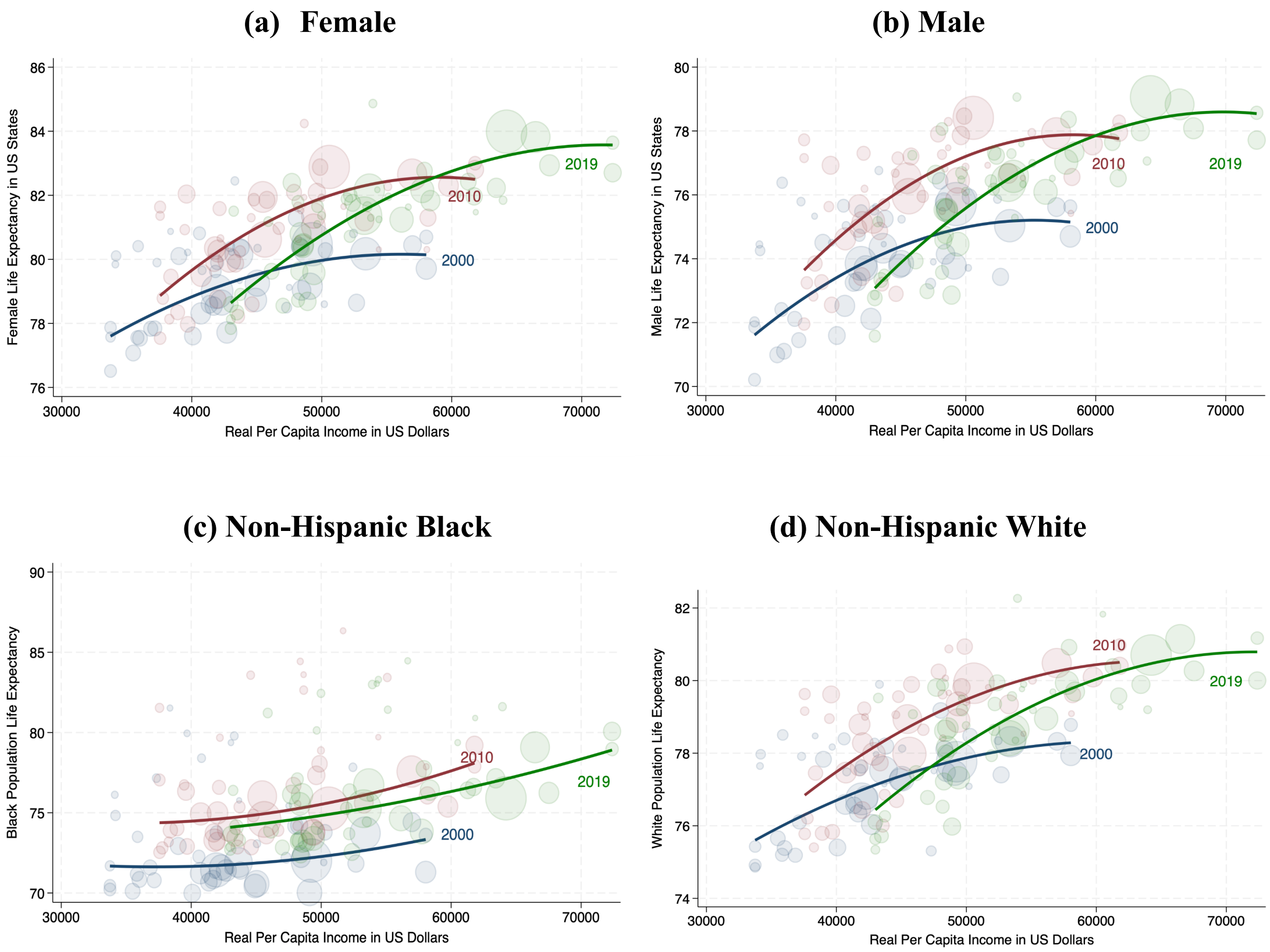

Notes: Preston Curves by sex and race. Temporal patterns in the evolution of the curves are similar across all race-gender groups. Of note, the convex curves for non-Hispanic Black Americans appear to be an artefact of population weighting and the uneven distribution of Black Americans in U.S. states. We do not reject the null of equivalence in the quadratic income term across the two racial groups. See Figure 1 for details on construction.

**Figure 4. Observed Life Expectancy Compared with Expected Life Expectancy Based on Economic, Social, Demographic, and Institutional Covariates**

**a) Observed LE in 2019 vs. LE Expected in 2019 based on 2010 Relationships Between LE and Covariates**

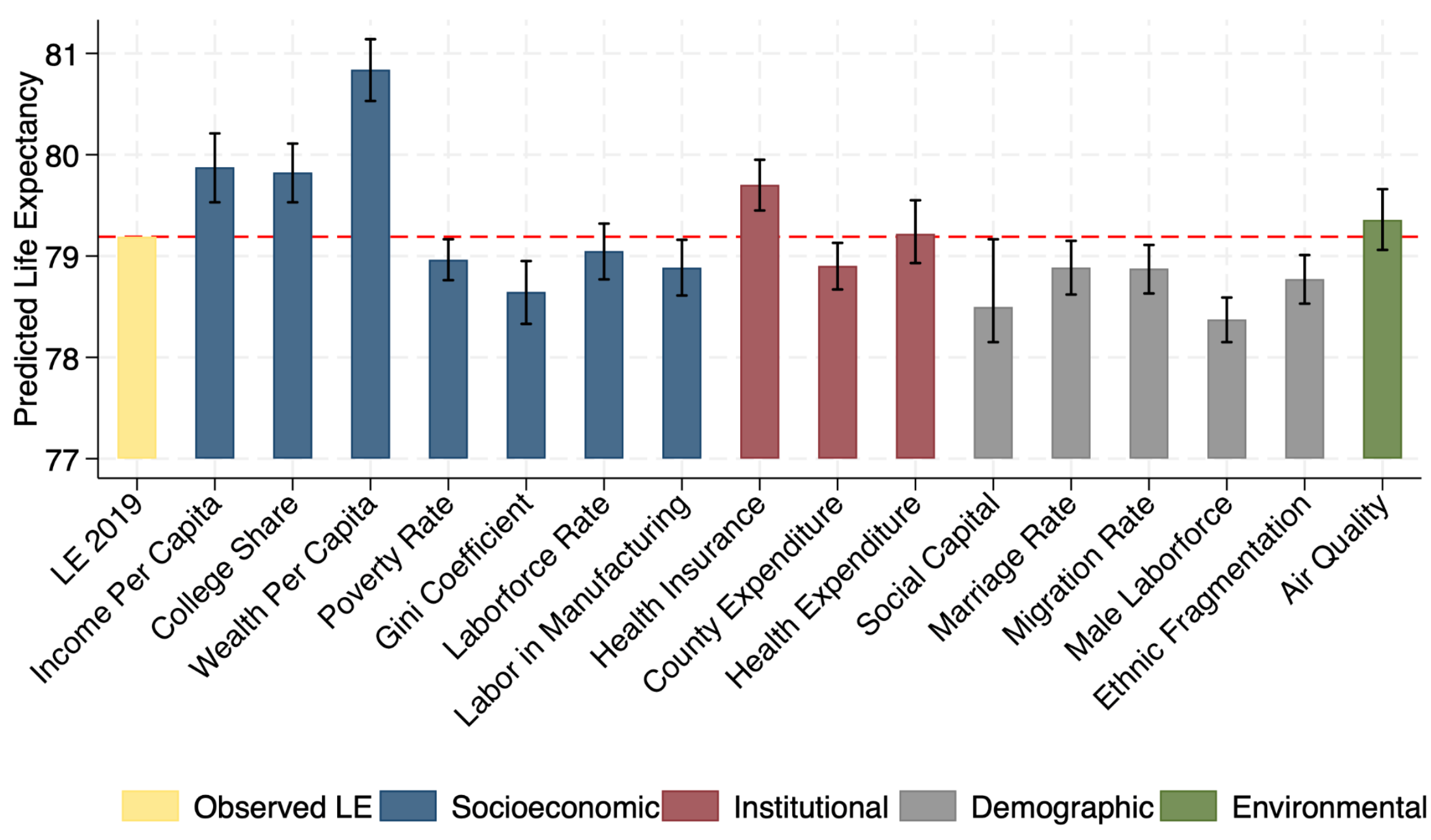

**b) Observed LE in 2010 vs. LE Expected in 2010 based on 2000 Relationships Between LE and Covariates**

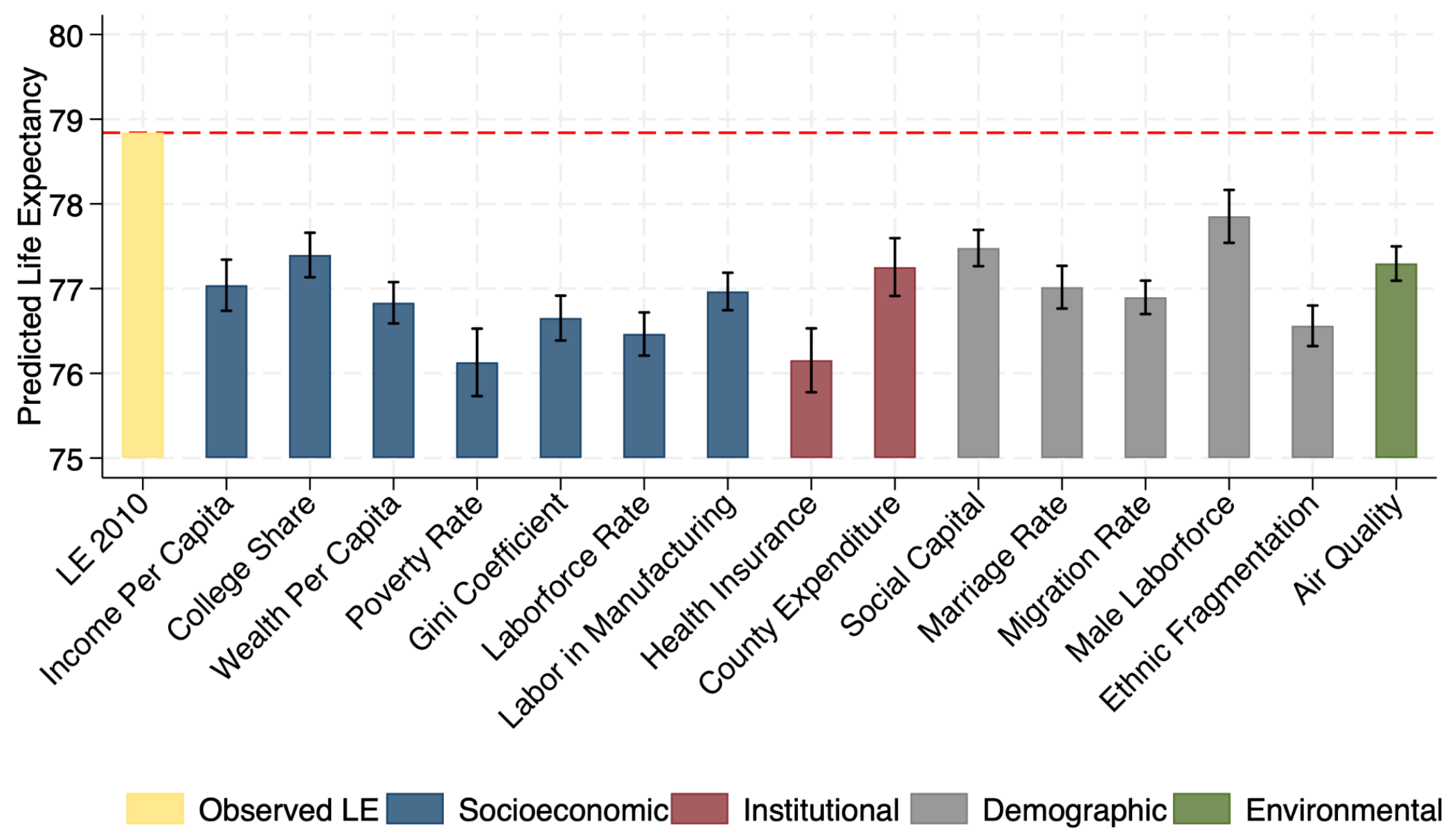

**c) Observed LE in 2019 vs. LE Expected in 2019 based on 2000 Relationships Between LE and Covariates**

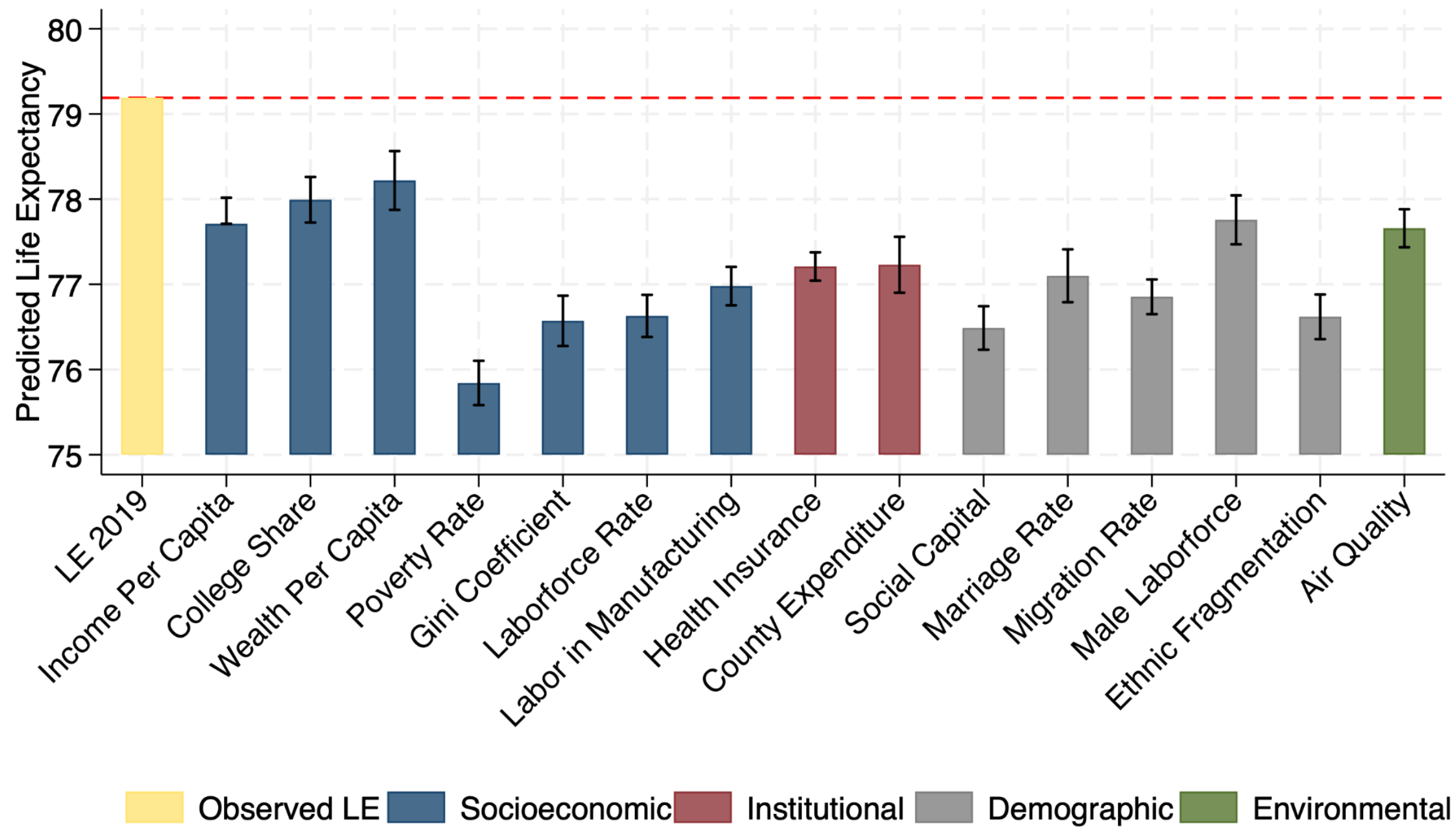

Notes: Results of a partial decomposition of life expectancy based on socioeconomic, institutional, demographic, and environmental covariates. We separately regress life expectancy and each covariate in the baseline year and use the resulting coefficients to estimate expected life expectancy using endline levels of covariates.

# Appendix

## Figure A1: All-Cause Mortality Preston Curves by Age Groups

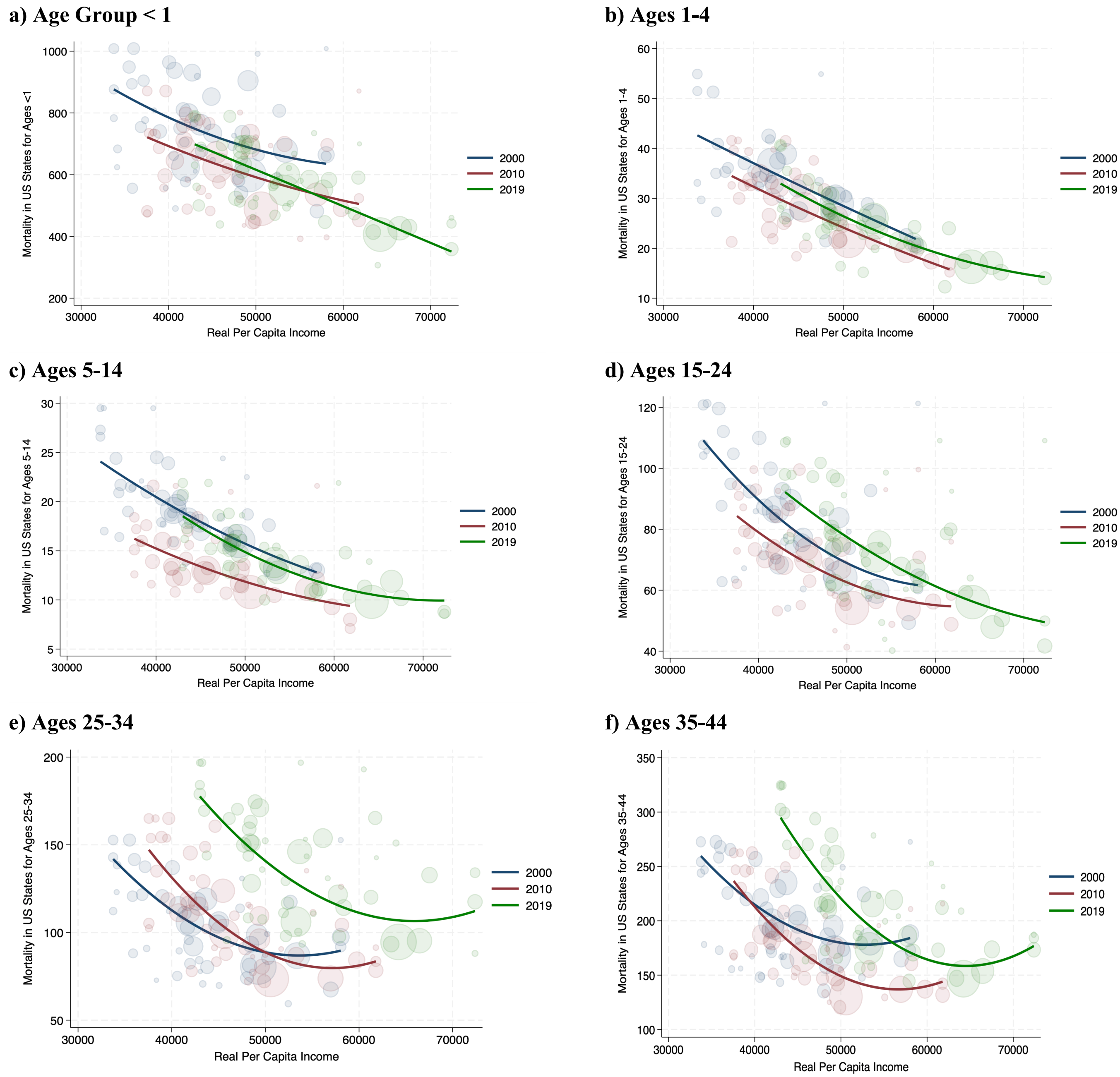

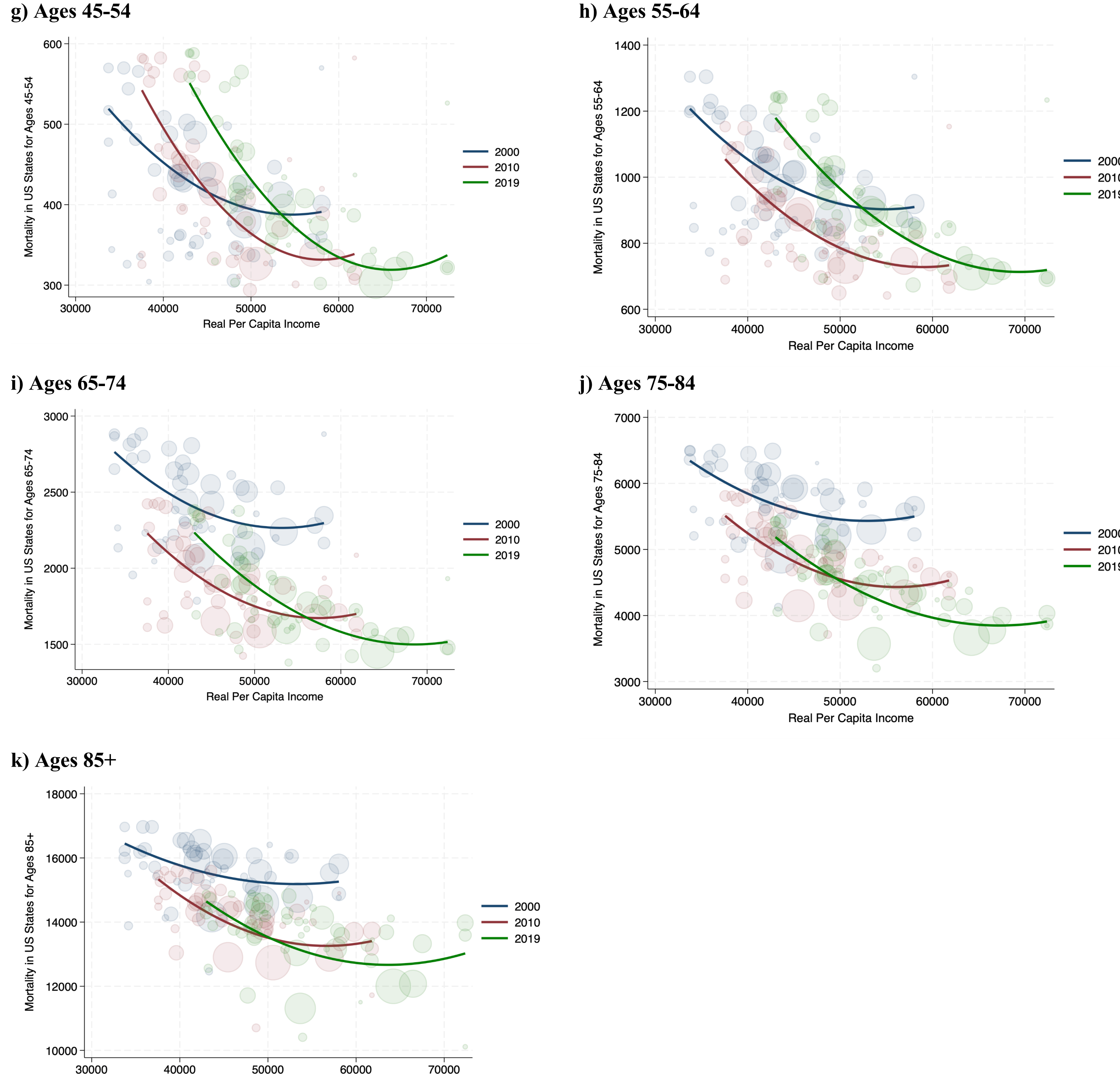

Notes: Shifts in Preston Curves between 2000-2019 using state-level, age-adjusted mortality rates. Older age groups (75+ years) followed the typical Preston curve pattern, with mortality decreasing across years (i.e., curve shifted down) as real per capita income increased (i.e., curve shifted right). Younger age groups (<64 years) violate the classic Preston curve pattern, with mortality increasing between 2010 and 2019 (i.e., curve stagnated or shifted up) as real per capita income increased (i.e., curve shifted right). For some age groups and income levels, mortality rates in 2019 exceeded those from 2000.

**Figure A2: Preston Curves with life expectancy plotted against logged income per capita**

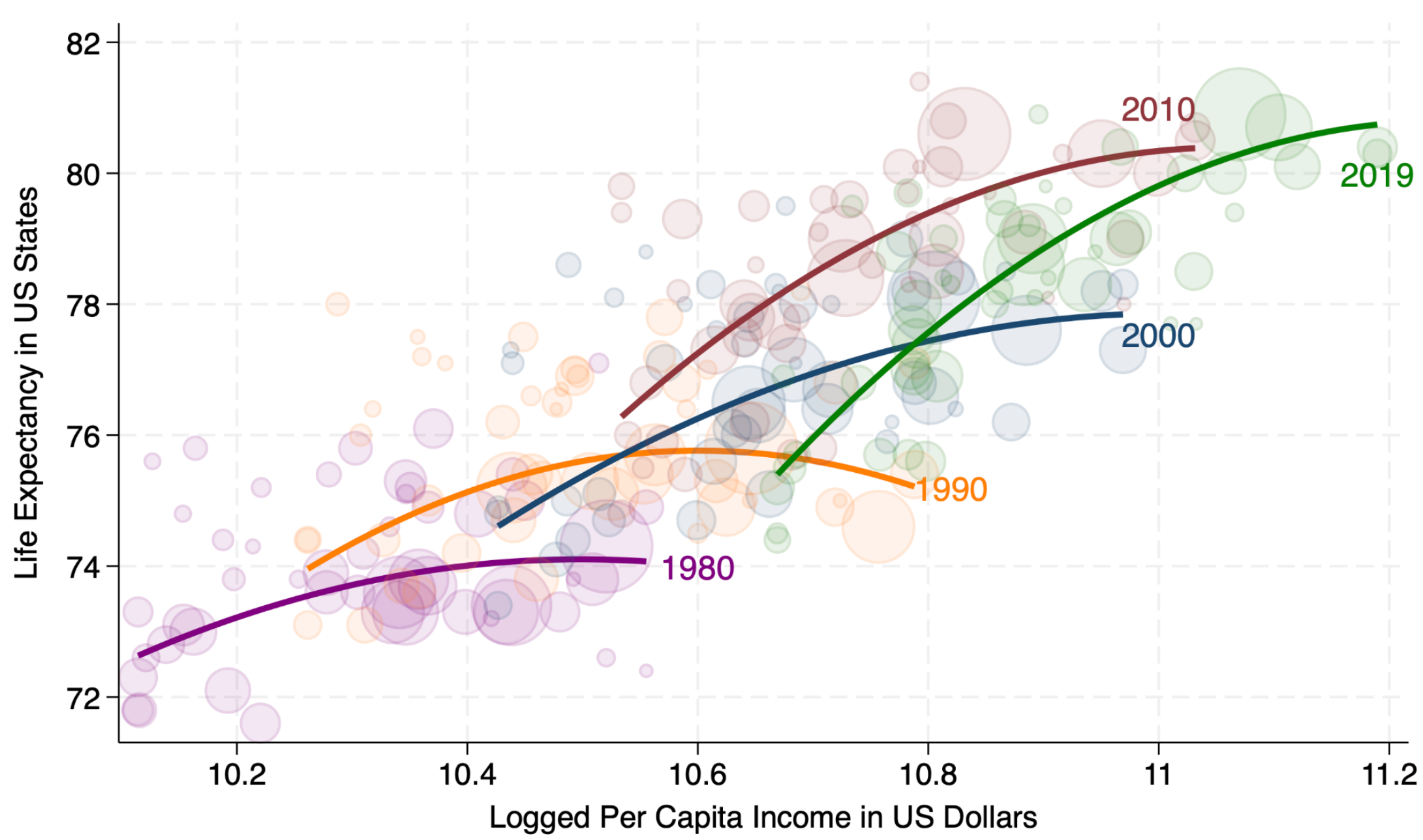

Notes: Preston curves using log of income per capita instead of income per capita levels. See Figure 1 for details on construction.

**Figure A3: Preston Curves with life expectancy against income per capita on a logged scale**

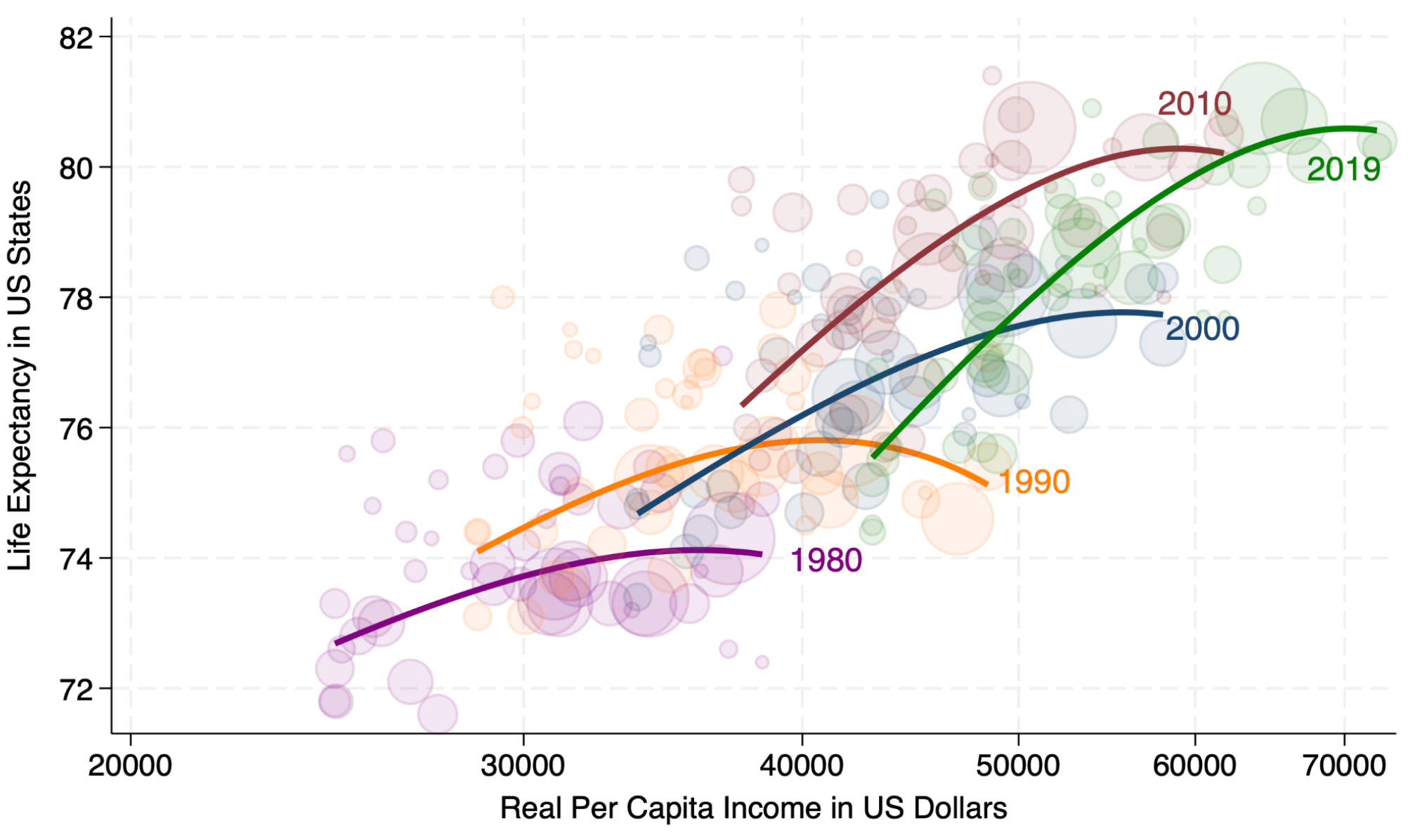

Notes: Preston Curves plotting income per capita on logged x-axis scale. See Figure 1 for details on construction.

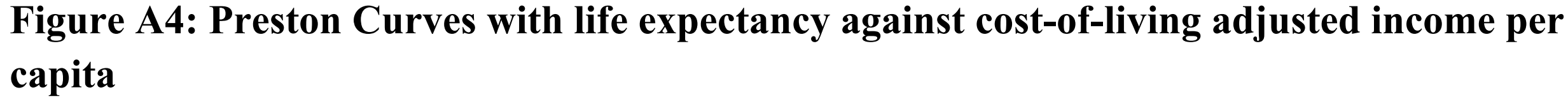

**Figure A4: Preston Curves with life expectancy against cost-of-living adjusted income per capita**

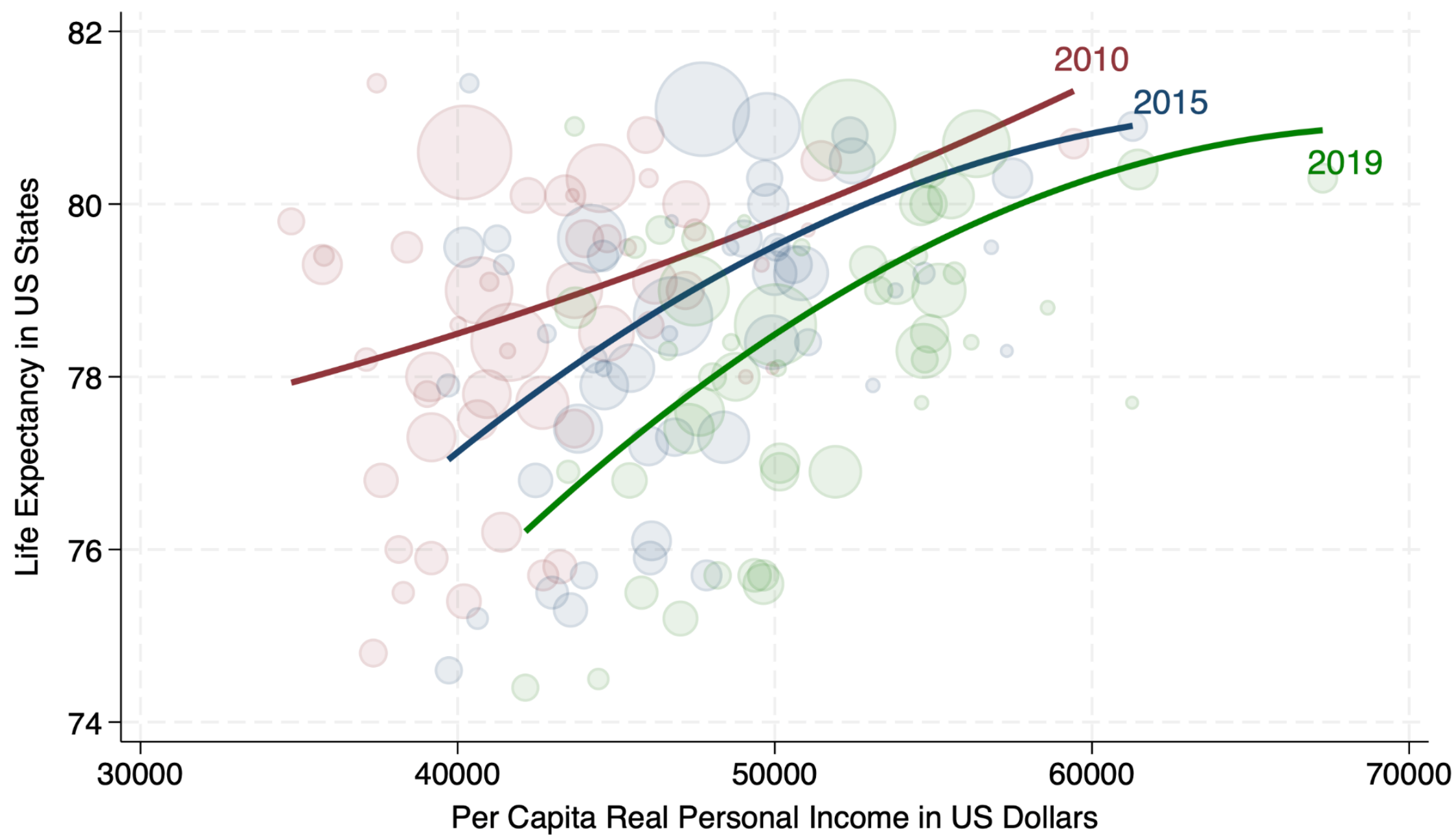

Notes: Preston curves constructed using BEA real per capita personal income (measured in chained 2012 dollars and available for 2010 onwards). This measure adjusts personal income for regional price differences using Regional Price Parities (RPPs) and for national inflation using the Personal Consumption Expenditures (PCE) price index. See Figure 1 for details on construction.